\documentclass[twocolumn,superscriptaddress,pre]{revtex4-1}

\newcommand{\nAB}{\ensuremath{\langle\boldsymbol n_A\cdot\boldsymbol n_B\rangle}}

\usepackage{amsmath} \usepackage{amssymb} \usepackage{graphicx}
\usepackage{hyperref}

\begin{document} \title{Folding pathways to crumpling in thermalized
elastic frames} \author{D.~Yllanes}\email{david.yllanes@czbiohub.org}\affiliation{Department of Physics and Soft
Matter Program, Syracuse University, Syracuse, NY, 13244} \affiliation{Chan Zuckerberg Biohub, San Francisco, CA 94158}\affiliation{Kavli
Institute for Theoretical Physics, University of California, Santa Barbara, CA
93106, USA} \affiliation{Instituto de Biocomputaci\'on y F\'{\i}sica de
Sistemas Complejos (BIFI), 50009 Zaragoza, Spain} 
 \author{D.R.~Nelson}\affiliation{Department
of Physics, Department of Molecular and Cellular Biology and School of
Engineering and Applied Sciences, Harvard University, Cambridge, MA 02138, USA}
\author{M.~J.~Bowick} \affiliation{Kavli Institute for Theoretical Physics,
University of California, Santa Barbara, CA 93106, USA} \date{\today}

\begin{abstract}
The mechanical properties of thermally excited two-dimensional crystalline membranes
can depend dramatically on their geometry and topology. A particularly relevant
example is the effect on the crumpling transition of holes
in the membrane. Here we use  molecular dynamics simulations to
 study the case of elastic frames (sheets with a single large hole
in the center) and find that the system approaches the crumpled phase
through a sequence of origami-like folds at decreasing length scales when temperature is 
increased. We use normal-normal correlation 
functions to quantify the temperature-dependent number of folds.
\end{abstract} \maketitle

\section{Introduction}\label{sec:intro}

Studies of two-dimensional materials such as
graphene~\cite{Katsnelson2012} have stimulated renewed exploration of the
statistical mechanics of elastic membranes. Thermalized elastic
membranes~\cite{Nelson2002,Nelson2004} have a low-temperature extended
(``flat'') phase with long-range order in the surface normals and a
scale-dependent bending rigidity ($\kappa$) and Young's
modulus ($Y$)~\cite{Nelson1987,Aronovitz1988,Guitter1989,
LeDoussal1992,Zhang1993,Bowick1996}. Without strong distant self-avoidance
these materials
are also believed to undergo a crumpling transition when the microscopic
bending rigidity is comparable to the scale of thermal
fluctuations~\cite{Bowick2001,Nelson2004, Kantor1987, Bowick1996,Cuerno2016}.
The crumpling transition, however, has never been convincingly observed in a
physical system. The challenge is to find a thin material that
exhibits both the flat phase and the crumpled phase.  One might expect that
soft flexible systems, such as pure amphiphilic bilayers~\cite{Shum2008}, would
then be the natural setting to observe the transition from a flat phase to a
crumpled phase. But here length scales become important. The flat phase itself
is stabilized by the strong thermal stiffening of the bending rigidity
resulting from soft flexural phonons that give rise to isotropic thermal
corrugations. These fluctuations set in only above the thermal length scale,
$\ell_\text{th}$, beyond which the renormalization of the bending rigidity becomes
comparable to the microscopic (bare) bending rigidity $\kappa_0$.  This scale, which
follows from the renormalization group flow of the bending rigidity,
is given by $\ell_\text{th} \sim \kappa_0/\sqrt{Y_0k_BT}$,
where $Y_0$ is the bare Young's modulus.
Even for the diblock copolymers of Ref.~\cite{Shum2008}, however,
this length is relatively large, $\ell_\text{th} \sim 300$ nm
at room temperature. It is therefore difficult to fabricate soft systems large enough to
harness these thermal effects. Graphene, on the other hand, has a thermal
length scale of the order of nanometers at room temperature, making it possible
to observe a crinkled flat phase~\cite{Nicholl2015,Blees2015}. The difficulty now is
accessing the crumpling transition which, considering that for graphene
$\kappa_0\approx1.25$ eV~\cite{Fasolino2007}, would
occur at temperatures of order $50\,000$ K!

\begin{figure}[b] \includegraphics[width=\linewidth]{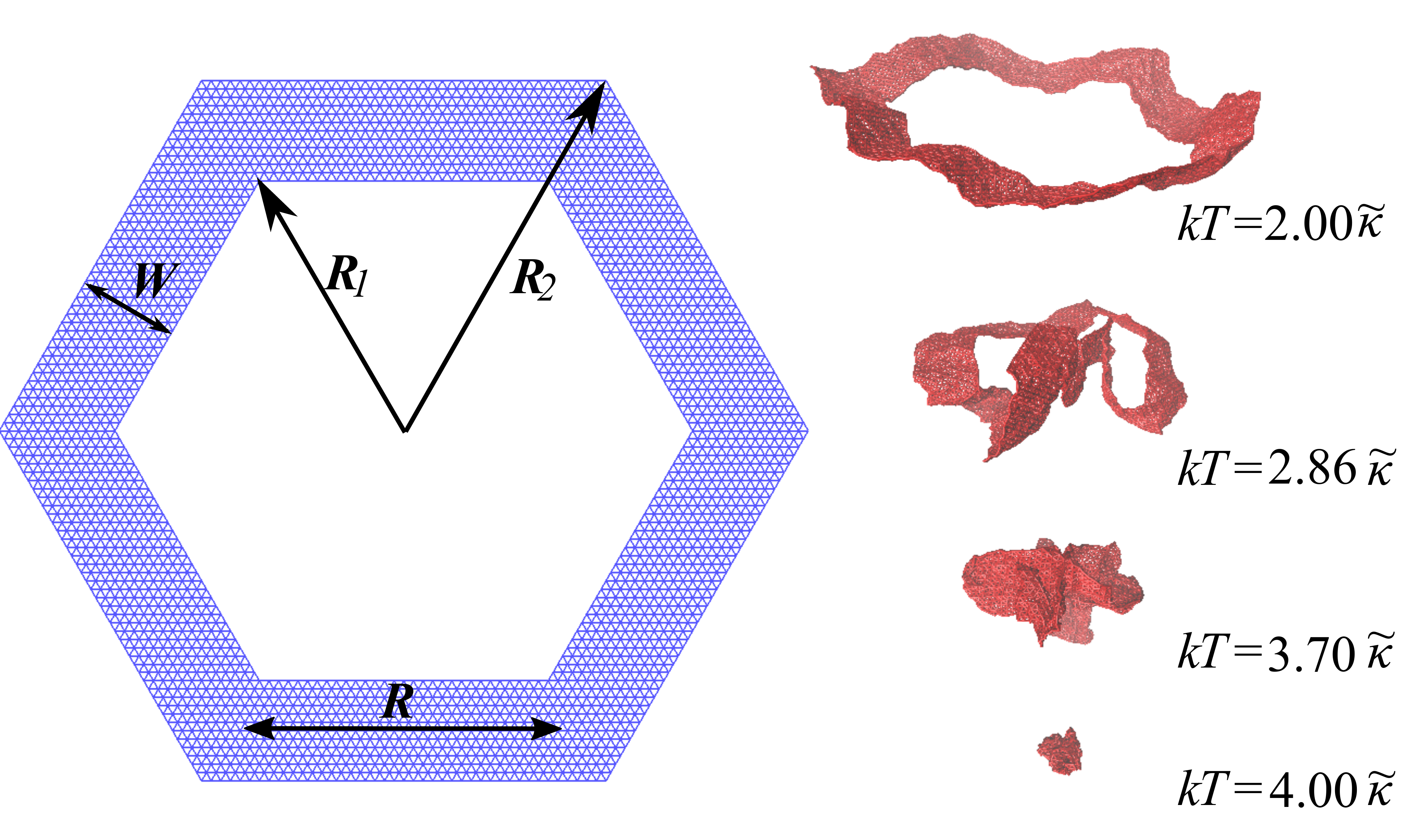}
\caption{Geometry of our system. We consider a
hexagonal frame of inner radius $R_1$ and outer radius $R_2=R_1+W$, whose
surface is triangulated.  Each bond in the resulting lattice contains a spring
potential, which characterizes the membrane's stretching, and each pair of
neighboring triangles has a cosine interaction potential depending on the angle between
their normals, characterized by a bending rigidity $\tilde\kappa$. See
\eqref{eq:H}.  As temperature is increased, the system
transitions from an extended phase, with long-range order in the normals, to a
crumpled phase.
 We show snapshots of the thermalized system at four
temperatures.All images in this figure have $R_1=30$ and $R_2=42$.
\label{fig:hexagono-crumpling}} \end{figure} 

In \cite{Yllanes2017} it was shown that the crumpling
transition can be dramatically lowered 
by either perforating the membranes with a periodic
array of holes or by excising a single large hole to form a thin frame. In both
cases there are two effects which lower the crumpling temperature. There is
less material to bend and there are fewer multiples of the thermal length scale
contributing to the stiffening of the renormalized bending rigidity. For
perforated membranes the growth is cut off by the mean spacing between holes
and for thin frames by the narrow frame width. Here we explore crumpling for a
thin hexagonal frame so as to preserve as many of the symmetries of the
triangular lattice as possible.  We find that crumpling is indeed enhanced,  but
also uncover a striking pathway to the crumpled phase. A well-known elementary
bending mode is an origami-like fold along a line \cite{Kantor1990,DiFrancesco1994,Santangelo2017}. 
The energy cost for such a fold
is proportional to the length of the fold line rather than the typical area
cost for introducing a complex network of crinkles designed to
crush a macroscopic piece of membrane \cite{Vliegenthart2006, Lahini2017}.

 With our molecular dynamics simulations we  find that crumpling occurs
via a sequence of these soft origami folds. The successive stages of crumpling
may be characterized by the distribution of the fold lines. The patterns
observed correspond to the fold lines for the exactly solvable model of planar
folding first studied in \cite{DiFrancesco1994}.

The rest of the paper is organized as follows: Section~\ref{sec:model} presents 
our coarse-grained crystalline model and describes our simulations. In 
Section~\ref{sec:order} we define an order parameter sensitive to the degree of
crumpling
based on normal-normal correlations and describe the folding process qualitatively.
This description is made more quantitative in Section~\ref{sec:corr} with the use
of the full normal-normal correlation function. Finally, Section~\ref{sec:geometry}
examines the effect of changing the size of the hole and Section~\ref{sec:conclusions}
summarizes our conclusions. We include an appendix that considers the normal-normal
correlation function in Fourier space.

\section{Model}\label{sec:model}

 We consider a  two-dimensional sheet of elastic material with a large hole
in the center (a \emph{frame}), which we model as a crystalline
membrane~\cite{Bowick2001}. In this representation, the sheet is discretized
with a tiling of equilateral triangles of side $a=1$, which defines a lattice
of unbreakable but elastic bonds. In order both to facilitate triangulation and to approach
a circular symmetry we use a hexagonal geometry with inner radius $R_1$ and
outer radius $R_2=R_1+W$ (Figure~\ref{fig:hexagono-crumpling}).   This system
can be regarded as six ribbons of 
length $R=(R_1+R_2)/2$ stitched together to form a hexagon.
On this
triangular lattice we define a standard coarse-grained elastic
Hamiltonian~\cite{Seung1988}:
 \begin{equation}\label{eq:H} \mathcal H =
\mathcal H_\text{stretch} +  \mathcal H_\text{bend}, \end{equation}
 where
\begin{align} \mathcal  H_\text{stretch} &= \frac12 \epsilon \sum_{\langle
i,j\rangle} (r_{ij}-a)^2, \\ \mathcal H_\text{bend} &= \tilde \kappa
\sum_{\langle \alpha,\beta\rangle} (1-\hat{\boldsymbol n}_\alpha\cdot
\hat{\boldsymbol n}_\beta) \label{eq:Hbend}\end{align} 
The sum in the stretching term is over
all pairs of lattice nearest neighbors $\langle i,j\rangle$. The bending
energy, on the other hand, depends on the angles between the normals
$\boldsymbol n_\alpha$, $\boldsymbol n_\beta$ of all pairs of triangles
$(\alpha,\beta)$ that share a side. Each of these pairs of triangles
defines a dihedral, which is how we will refer to each term
in the bending Hamiltonian in the rest of the paper.

The coupling constants $\epsilon$ and $\tilde\kappa$ are directly related to
the Young's modulus and bare bending rigidity of continuum elastic theory
($Y_0=2\epsilon/\sqrt3$, $\kappa_0=\sqrt3\tilde\kappa/2$ \cite{Seung1988}).  As temperature
increases, the frame transitions from a thermally excited flat phase, with long-range order in the
normals, to a crumpled phase (see Figure~\ref{fig:hexagono-crumpling}).  This
crumpling transition has already been thoroughly studied for unperforated
sheets~\cite{Bowick2001,Nelson2004, Kantor1987,LeDoussal1992, Bowick1996,
Cuerno2016} as well as for sheets with
perforations~\cite{Yllanes2017}. In the following, therefore, we do not
focus on the crumpling transition itself, but instead study 
an intermediate regime, where we show that crumpling is 
realized through a series of folding pathways.

Aside from an overall
constant factor that defines the energy units, what differentiates
one elastic material from another is the ratio $\epsilon/\tilde\kappa$.
Following \cite{Bowick2017,Yllanes2017} we set $\epsilon=1440\tilde \kappa/a^2$, which 
approximates graphene values without making the system too rigid and hard to
thermalize. In practice, this choice makes very little difference
in the final results, since the crumpling point depends only logarithmically
on the Young's modulus (see, e.g.,~\cite{Kosmrlj2016} and compare the 
crumpling temperatures in~\cite{Cuerno2016} and~\cite{Yllanes2017}).

We have carried out Molecular Dynamics simulations of model~\eqref{eq:H} for different
system sizes and a wide temperature range (working in an $NVT$ ensemble 
with a standard Nos\'e-Hoover thermostat~\cite{Nose1984,Hoover1985}). Our 
simulations were implemented in the HOOMD-blue package~\cite{Glaser2015,Anderson2008}
and run on Tesla GPUs. Following~\cite{Yllanes2017}, we use
a timestep of $\Delta t=0.0025\tau$, where $\tau=\sqrt{ma^2/kT}$ is
the Lennard-Jones unit of time (expressed in terms of the particle mass
$m$) and we use natural units with $a=m=1$.
All energies are measured in units of $kT$. For each system size
and temperature we run for $10^9$ MD steps and use a jackknife
procedure~\cite{Amit2005} to estimate statistical errors. For
our larger frames with $R_1=70, R_2=82$ these runs take from 60
to 70 hours on a Tesla K40m GPU.
\begin{figure}[t]
\includegraphics[height=\linewidth,angle=270]{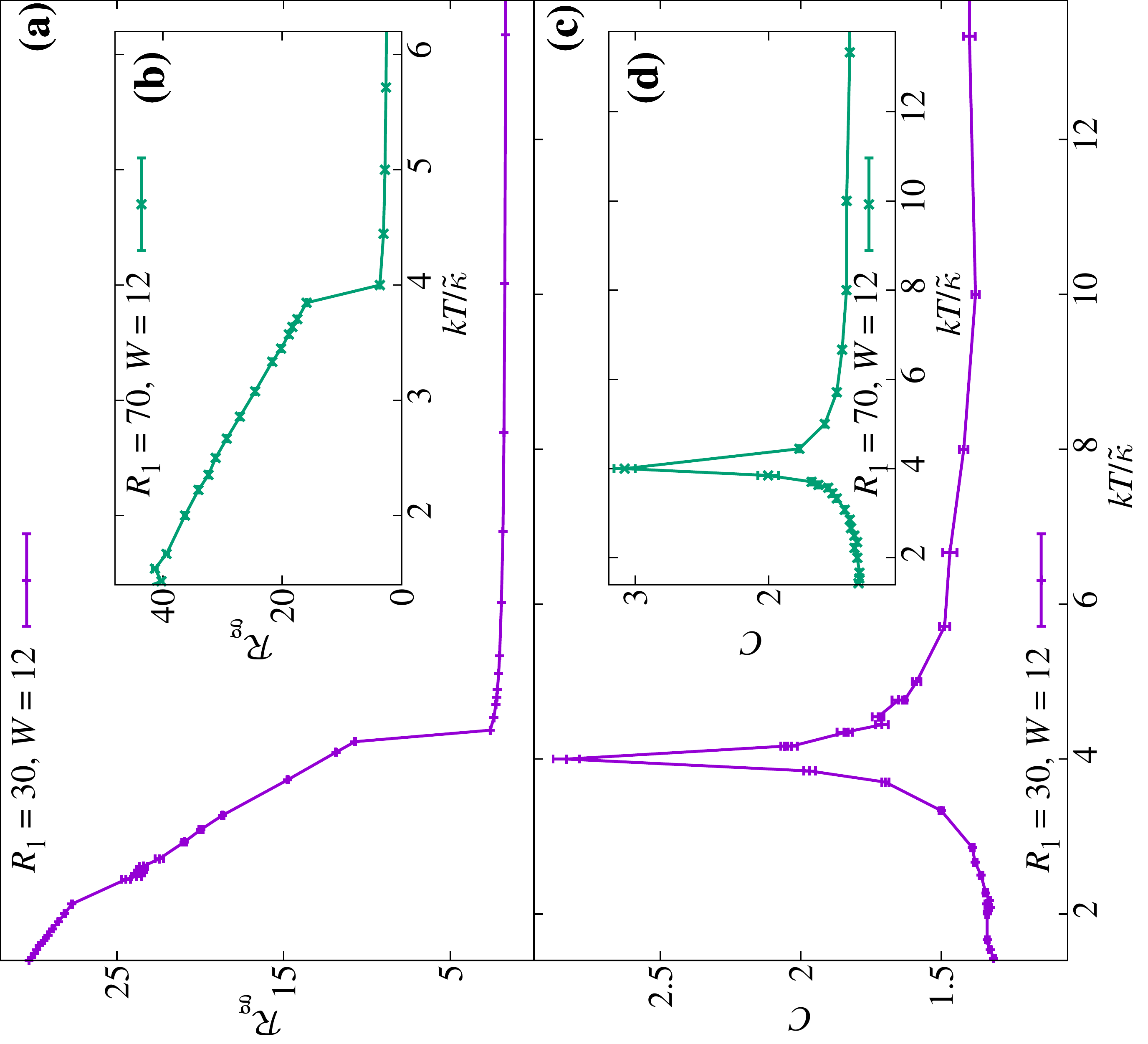}
\caption{The crumpling transition of a frame. 
We plot the radius of gyration, Eq. \eqref{eq:Rg}, as a function of
temperature for a frame of $R_1=30$ and $R_2=42$ ($W=12$)
in panel (a). Panel (b) shows the corresponding
curve for a hexagon of $R_1=70$ and the same width $W=12$.
For the same systems, we plot
the specific heat $C$ in panels (c) and (d). This 
quantity has a sharp peak at the transition point, 
$k T_\text{c}/\tilde\kappa \approx4.00$  
for both sizes.
\label{fig:Rg}}
\end{figure}
\begin{figure}
\centering
\includegraphics[width=.6\linewidth]{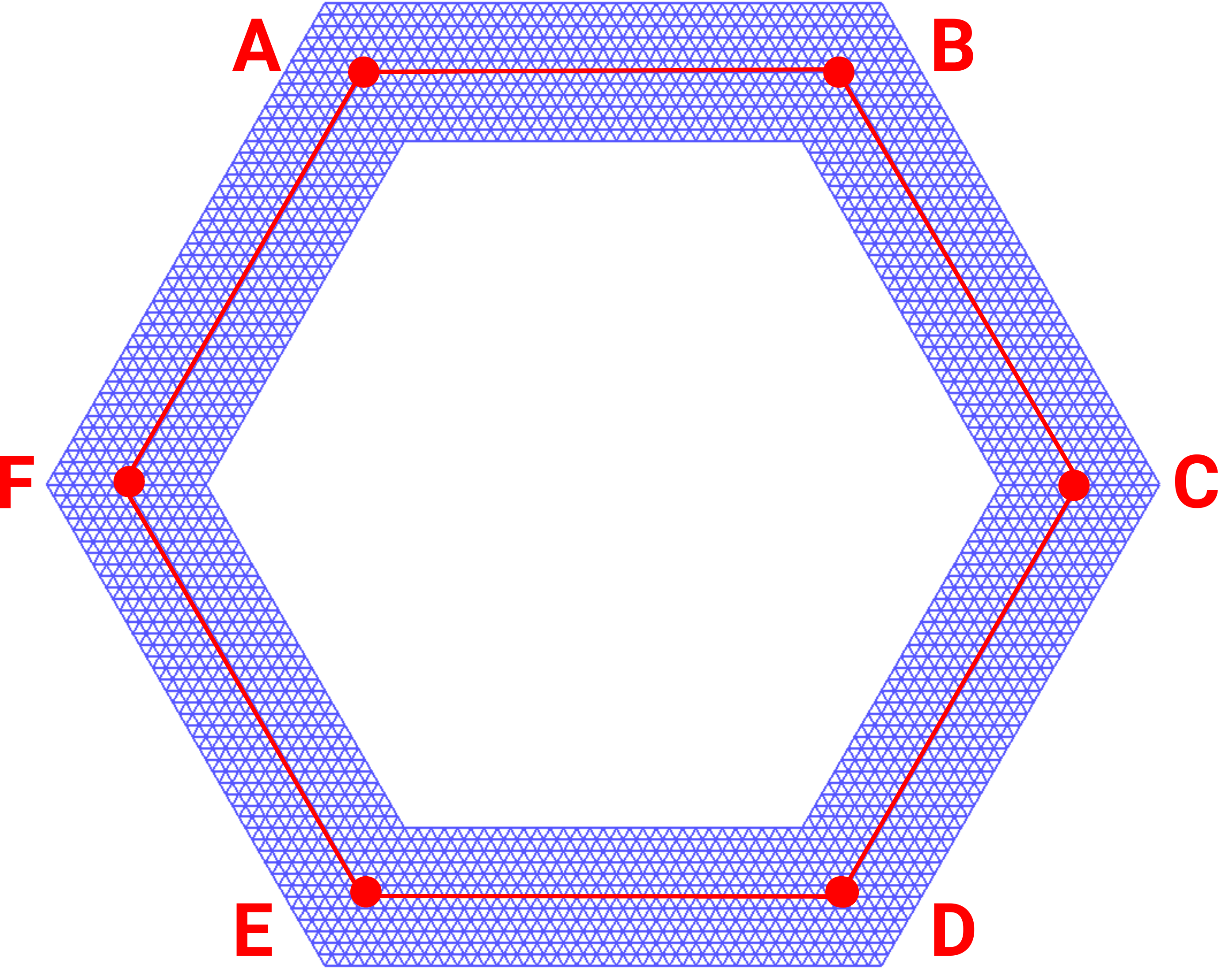}

\includegraphics[height=\linewidth,angle=270]{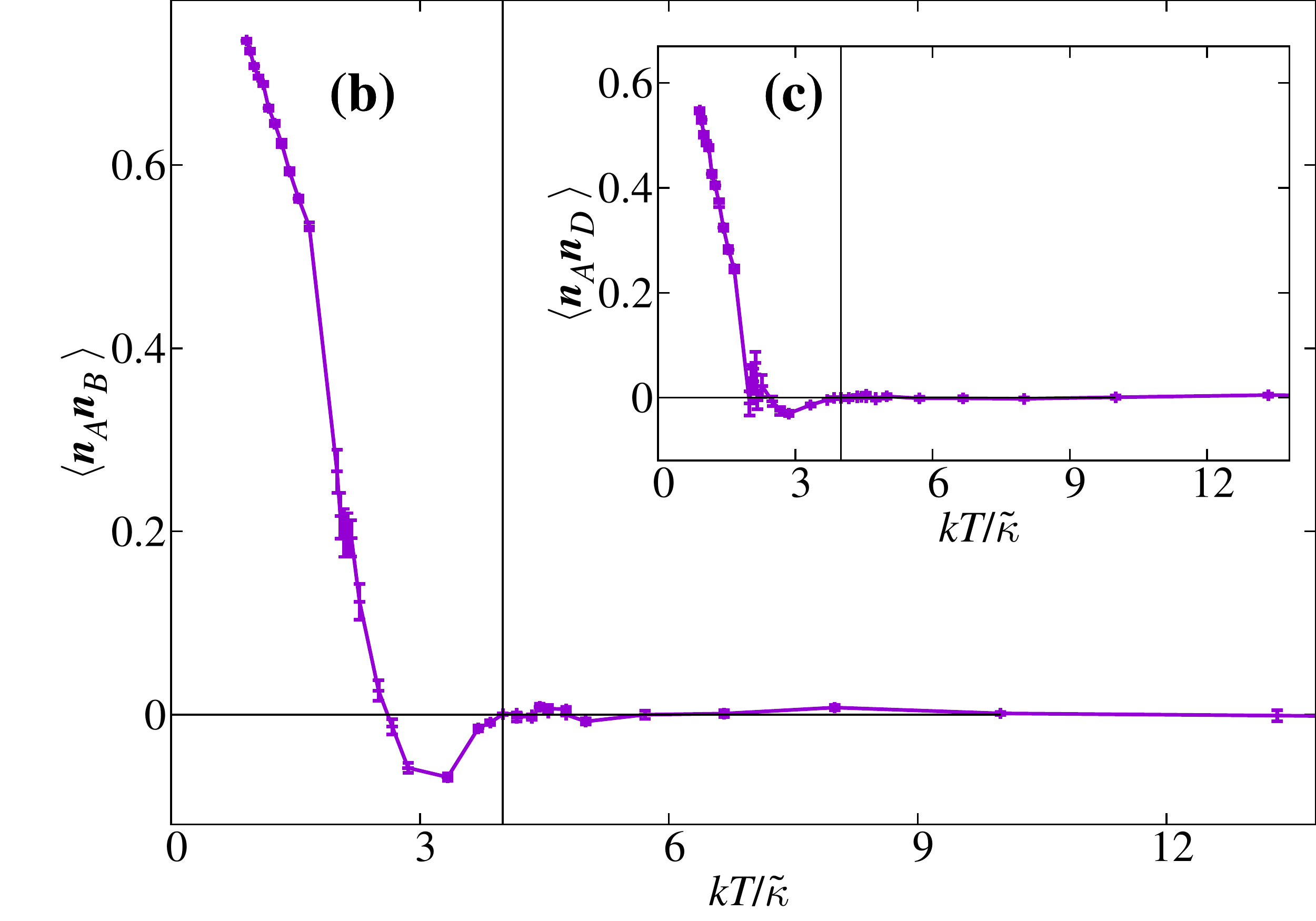}
\caption{We will consider the correlation between the red points 
along the central hexagon of radius $R=(R_1+R_2)/2$ (a).
In panel (b) we plot of the correlation $\langle \boldsymbol n_A\cdot
\boldsymbol n_B\rangle$
between normals at opposing vertices along the same side of the hexagon (we
consider the average
of the six equivalent pairs of points). This correlation goes to zero
well before the critical point $kT_\text{c}\approx 4.00\tilde \kappa$ identified 
in Figure~\ref{fig:Rg} (marked with a vertical line) and in fact 
becomes negative, indicating an anticorrelation between normals. 
Inset (c)
shows the corresponding graph for diagonally opposed vertices,
$\langle\boldsymbol n_A\cdot \boldsymbol n_D\rangle$
(and two equivalent pairs), which goes to zero even earlier but is much noisier.
\label{fig:norm}}
\end{figure}
\begin{figure*}[t]
\includegraphics[width=\linewidth]{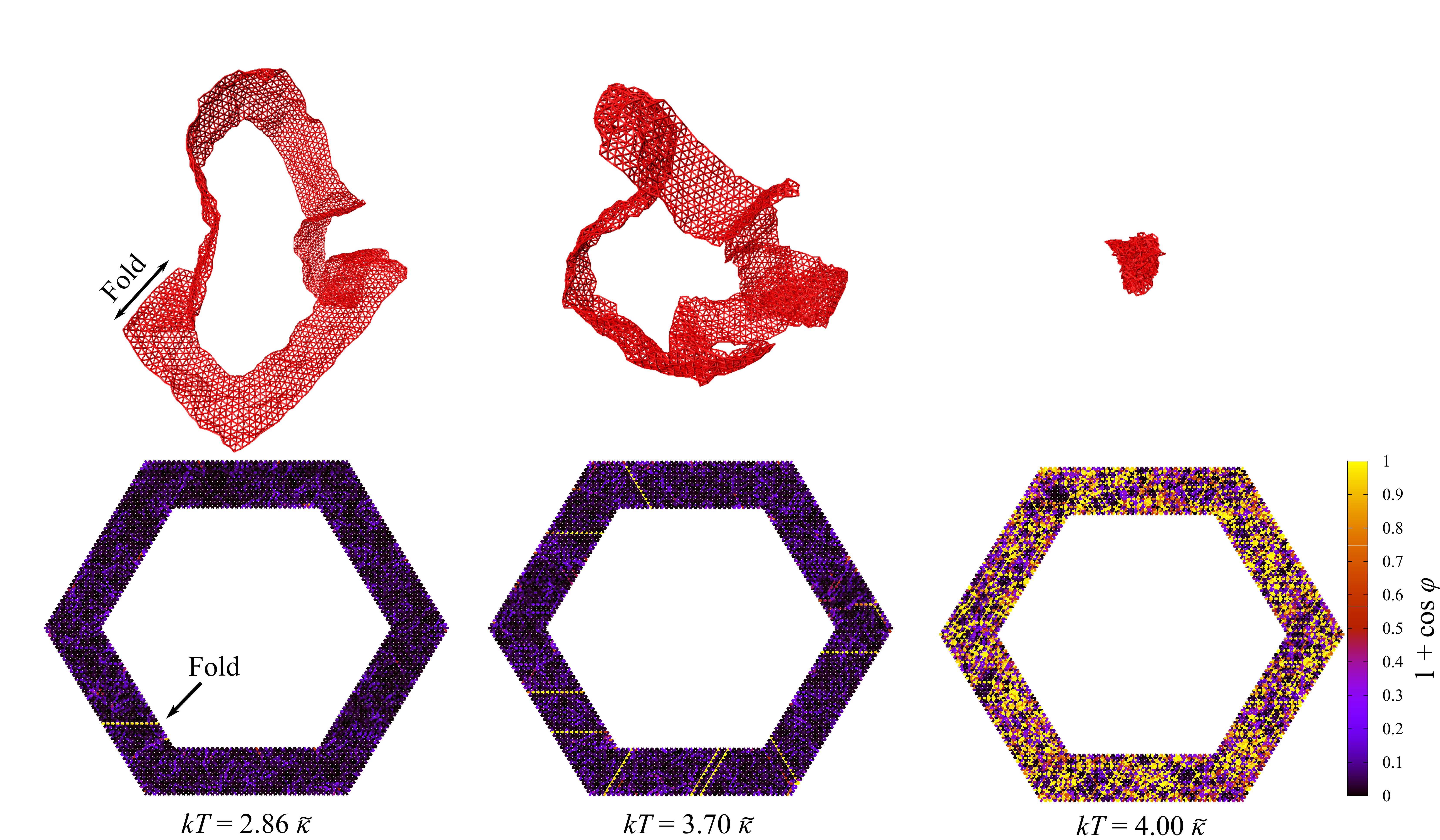}
\caption{Heat map of bending energy on a hexagonal frame
with various degrees of crumpling. For three
temperatures we show two representations of a typical thermalized configuration.
On the top row we simply show the three-dimensional state of the sheet. On the bottom 
row, we plot via a heat map the value of $(1+\cos\varphi)$ for all the dihedrals in the
system, where $\varphi$ is the angle between normals ($\varphi=\pi$ for a flat
dihedral, where the energy is minimal). This quantity is a measure of 
the bending energy~\eqref{eq:Hbend} in natural units. For the lowest
temperature shown, $kT=2.86\tilde\kappa$, most of the energy is 
concentrated in a single sharp fold.  As we approach the critical point of
$kT_\text{c} \approx 4.0\kappa$, an increasing number of 
well-defined radial folds appear. 
\label{fig:heat}}
\end{figure*}

\section{The crumpling order parameter}\label{sec:order}

The standard method to study the crumpling transition itself is to 
consider the radius of gyration of the frame
 \begin{align}\label{eq:Rg} \mathcal R_\text{g}^2
&=\frac{1}{3N} \sum_{i=1}^N \langle\boldsymbol R_i\cdot \boldsymbol
R_i\rangle,& \boldsymbol R_i = \boldsymbol r_i - \boldsymbol r_\text{center of mass}.
\end{align}
In this equation and in the following, $\langle\cdot\rangle$ will denote
a thermal average. At low temperatures, $\mathcal R_\text{g}^2 \sim R_2^{2}$, 
while in the crumpled phase $\mathcal R_\text{g}^2 \sim \log(R_2/a)$. Panels
(a) and (b) of Figure~\ref{fig:Rg} show $\mathcal R_\text{g}$ as a function
of temperature for two frame sizes.

A very clear signal of the phase transition can also be obtained 
by plotting the specific heat $C$ of our elastic frames
a function of temperature, which can be computed
with a fluctuation-response formula~\cite{Yllanes2017}:
 \begin{equation}\label{eq:C} C
= \frac1N \bigl( \langle \mathcal H^2\rangle - \langle \mathcal
H\rangle^2\bigr)\ .  \end{equation}
This quantity has a very sharp peak at the transition point
of $kT_\text{c}\approx4.0 \tilde\kappa$ for
both values of $R_1$ studied (see Figures~\ref{fig:Rg}c and d).

Finite-size scaling studies of the crumpling transition
have already been done in considerable detail both for pristine~\cite{Cuerno2016} and
perforated~\cite{Yllanes2017} sheets. Here, we are interested
in following the detailed geometry of the frame as the transition
is approached. To this end, we have found that 
a system size of $R_1=30$ and $R_2=42$ provides a good compromise:
The frame is large enough for finite-size effects to be negligible
(compare the position of the peaks for $R_1=30$ and $R_1=70$ in Figure~\ref{fig:Rg})
and to provide good statistics, yet small enough for us to follow 
individual folds and creases in snapshots. In what follows we shall
always consider this particular system size.

Even though $\mathcal R_\text{g}$ is a useful proxy for a
crumpling order parameter, it is illuminating to
study normal-normal correlations directly. 
Because our frames are made up of six ribbons 
of length $R=(R_1+R_2)/2$ stitched 
together to form a hexagon (see Fig. \ref{fig:hexagono-crumpling})
we first consider a long elastic ribbon at a low temperature. 
Then the correlation between the normal
at the beginning ($O$) and end ($X$) of the ribbon 
is~\cite{Kosmrlj2016}  
\begin{equation}\label{eq:normals}
\langle\hat{\boldsymbol n}_O\cdot \hat {\boldsymbol n}_X\rangle_L = 1 - \frac{kT}{2\pi\kappa_0}
\left[ \eta^{-1} + \log\left(\frac{\ell_\text{th}}{a}\right) + \mathcal C \frac{kT}{\kappa_0} \left(\frac{\ell_\text{th}}{L}\right)^\eta\right]\, .
\end{equation}
In this equation, $\eta\approx0.8$ is a critical exponent, $\mathcal C$ is a positive constant
of order unity and $\ell_\text{th}^2 = 2\sqrt{3}\pi^3 \tilde\kappa^2 / 3 kT \epsilon$ 
is the thermal length scale. For sufficiently small $kT$, this correlation 
is positive even for $r_{OX}=L\to\infty$. Conversely, we could in principle
 approximate the crumpling transition as the point where $\langle \boldsymbol n_O\cdot \boldsymbol n_X\rangle\approx0$
for large $r_{OX}=L$.

This test is easily done by considering our frame to be a ribbon of width $R_2-R_1=W$, with periodic boundary conditions
in the azimuthal direction. 
If we then consider a central hexagon of radius $R=(R_1+R_2)/2$,
we can identify $\langle \boldsymbol n_O\cdot \boldsymbol n_X\rangle$ with $\langle \boldsymbol n_A\cdot \boldsymbol n_D\rangle$,
 the correlation between diagonally opposed vertices (see Figure~\ref{fig:norm}) \footnote{We 
actually consider the average of {$\langle \boldsymbol n_A\cdot \boldsymbol n_D\rangle$, $\langle \boldsymbol n_B\cdot \boldsymbol n_E\rangle$
and $\langle \boldsymbol n_C\cdot \boldsymbol n_F\rangle$.}}.

If we plot $\langle \boldsymbol n_A\cdot \boldsymbol n_D\rangle$ as a
function of temperature, however, we find that it goes to zero well before the crumpling
temperature of $kT_\text{c}\approx 4.0\tilde \kappa$, although the data is
rather noisy. We can obtain a better plot by studying instead the correlation
between opposing vertices along the same side of the hexagon ($\langle
\boldsymbol n_A\cdot  \boldsymbol n_B\rangle$ and five symmetric equivalents). This
quantity goes to zero at a higher temperature than $\langle \boldsymbol
n_A\cdot \boldsymbol n_D\rangle$, but still well before the actual crumpling
transition. Furthermore, for a finite temperature range, $\langle\boldsymbol
n_A\cdot \boldsymbol n_B\rangle$ is negative, indicating anticorrelation
between normals. 
\begin{figure}
\includegraphics[height=\linewidth,angle=270]{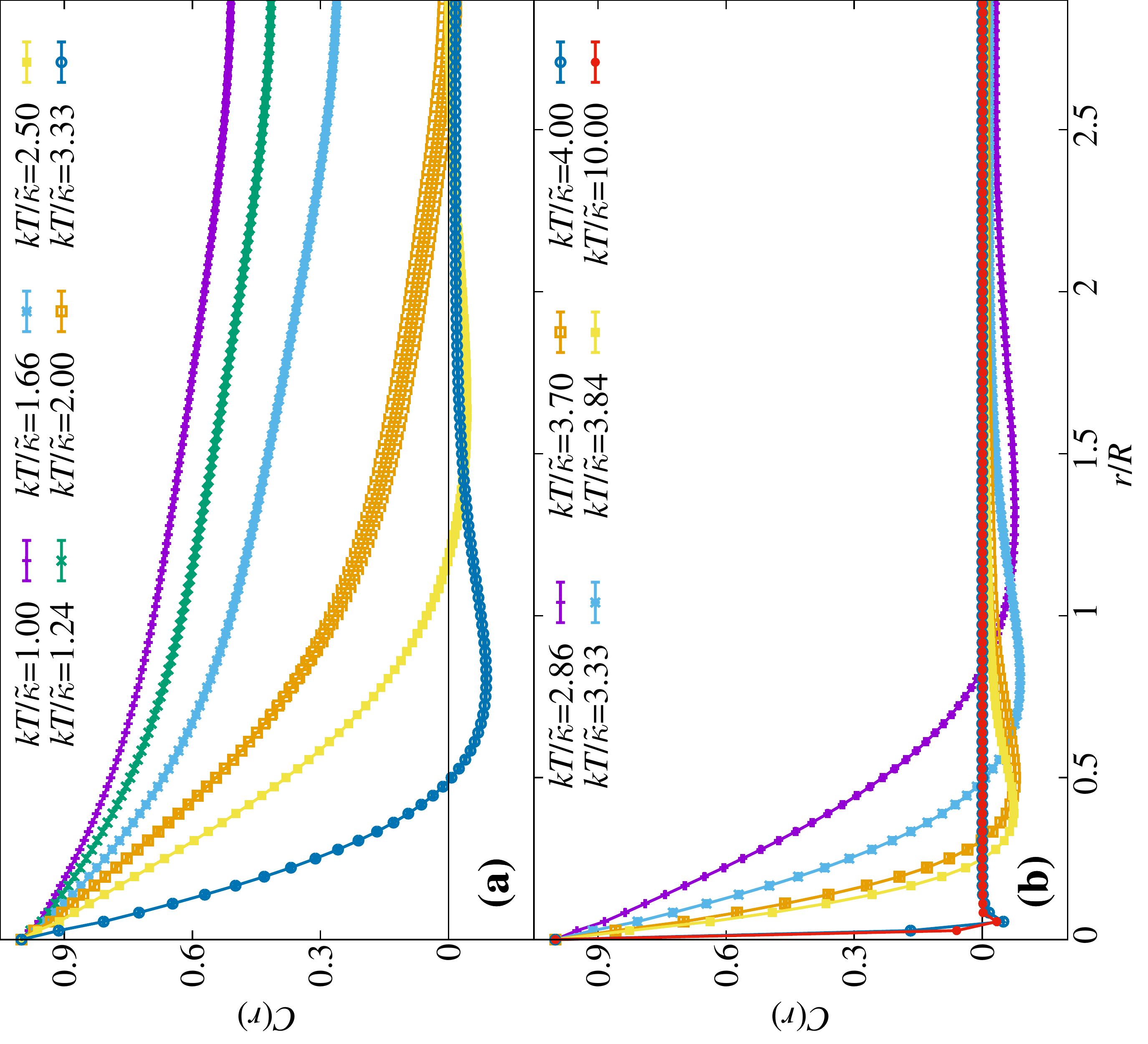}
\caption{Normal-normal correlation as a function of the arc length $r$ 
along the central hexagon of Figure \ref{fig:norm} for many temperatures
from the flat into the crumpled phase ($R_1=30, W=12$ and hence $R=36$). 
For low temperatures, the correlation function exhibits long-range order and
does not go to 
zero at long distances. As we approach the critical point of $kT_\text{c}
\approx 4.00~\tilde\kappa$, the correlation function
goes to zero for finite $r$ and reaches a well-defined negative minimum,
indicating the typical length scale of the folds in the system. At $kT \geq
4.00\tilde\kappa$, the folds occur at a microscopic scale, so the only anticorrelation
happens for the distance given by our discretization (i.e., between vertices at
opposing ends of the same dihedral). 
\label{fig:corr}}
\end{figure}

\begin{figure}
\includegraphics[height=\linewidth,angle=270]{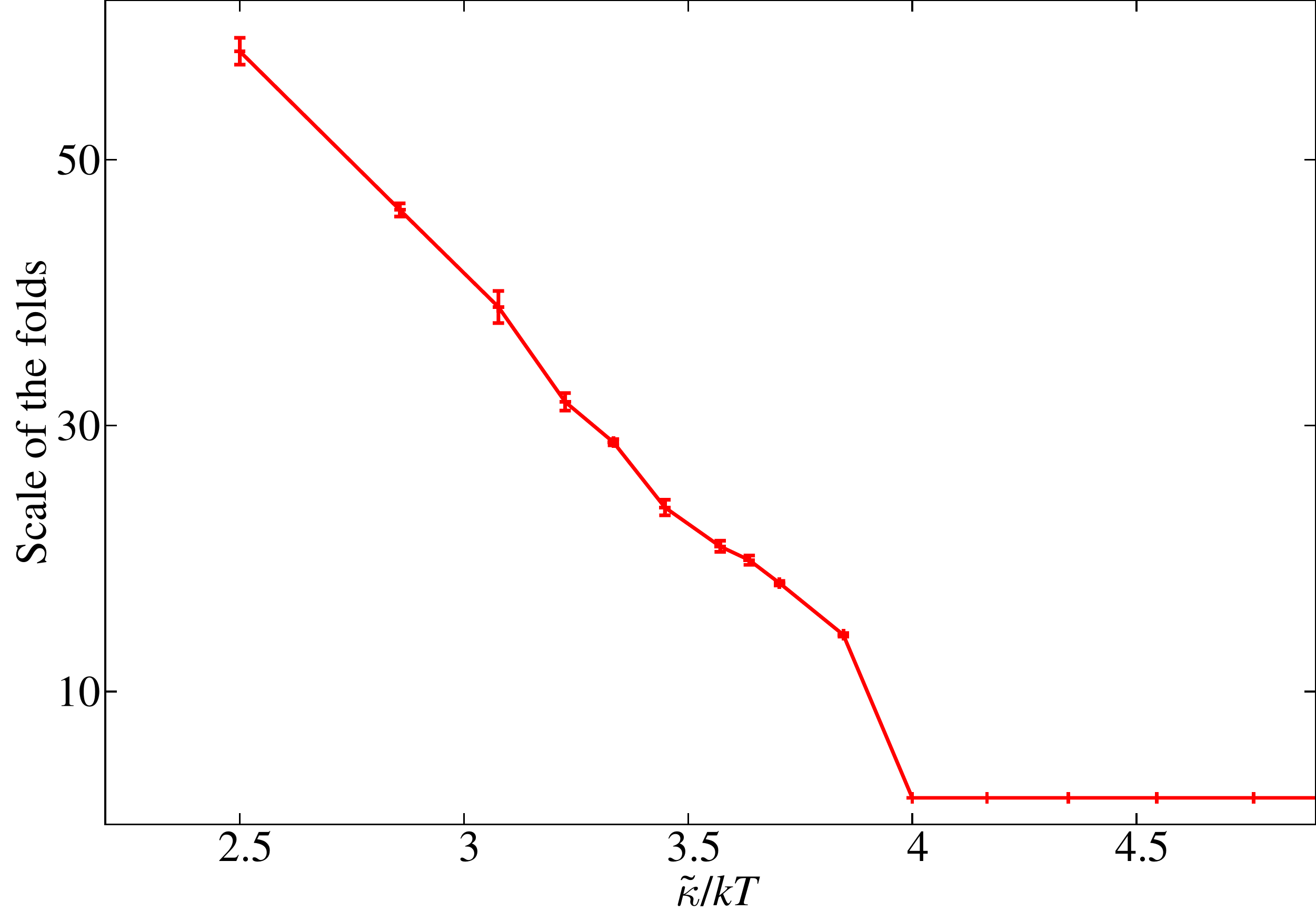}
\caption{Scale of the folds as a function of temperature. We 
plot the position of the minimum of $C(r)$ in (Figure~\ref{fig:corr})
for each temperature, showing how the distance between folds 
decreases as we approach the crumpling point. For $kT\lesssim2.4\tilde\kappa$
the minimum disappears and $C(r)$ becomes monotonic.
\label{fig:scale}}
\end{figure}

Let us consider the meaning of this negative correlation.  
The easiest way to achieve $\boldsymbol n_A \cdot \boldsymbol n_B=-1$ 
is to create a single fold between the two points. Of course, our 
frame is fluctuating thermally. A measurable negative
ensemble averaged $\langle \boldsymbol n_A \cdot \boldsymbol n_B\rangle$, 
however, suggests that typical instantaneous configurations of the frame
might have well defined origami folds, separated by a distance
comparable to $|\boldsymbol r_A-\boldsymbol r_B|$, but whose location
varies with time. Upon recalling that the correlation for 
the even more distant points such as $A$ and $D$ goes to zero at an even lower
temperature [Figure~\ref{fig:norm}-(c)], we can hypothesize
that there exists an intermediate regime before crumpling
where the system concentrates its bending energy on a small number 
of well-defined folds. We expect that the distance between single folds
is a function of temperature, which becomes quite small as $T$ approaches
the crumpling transition.
\begin{figure}
\includegraphics[height=\linewidth,angle=270]{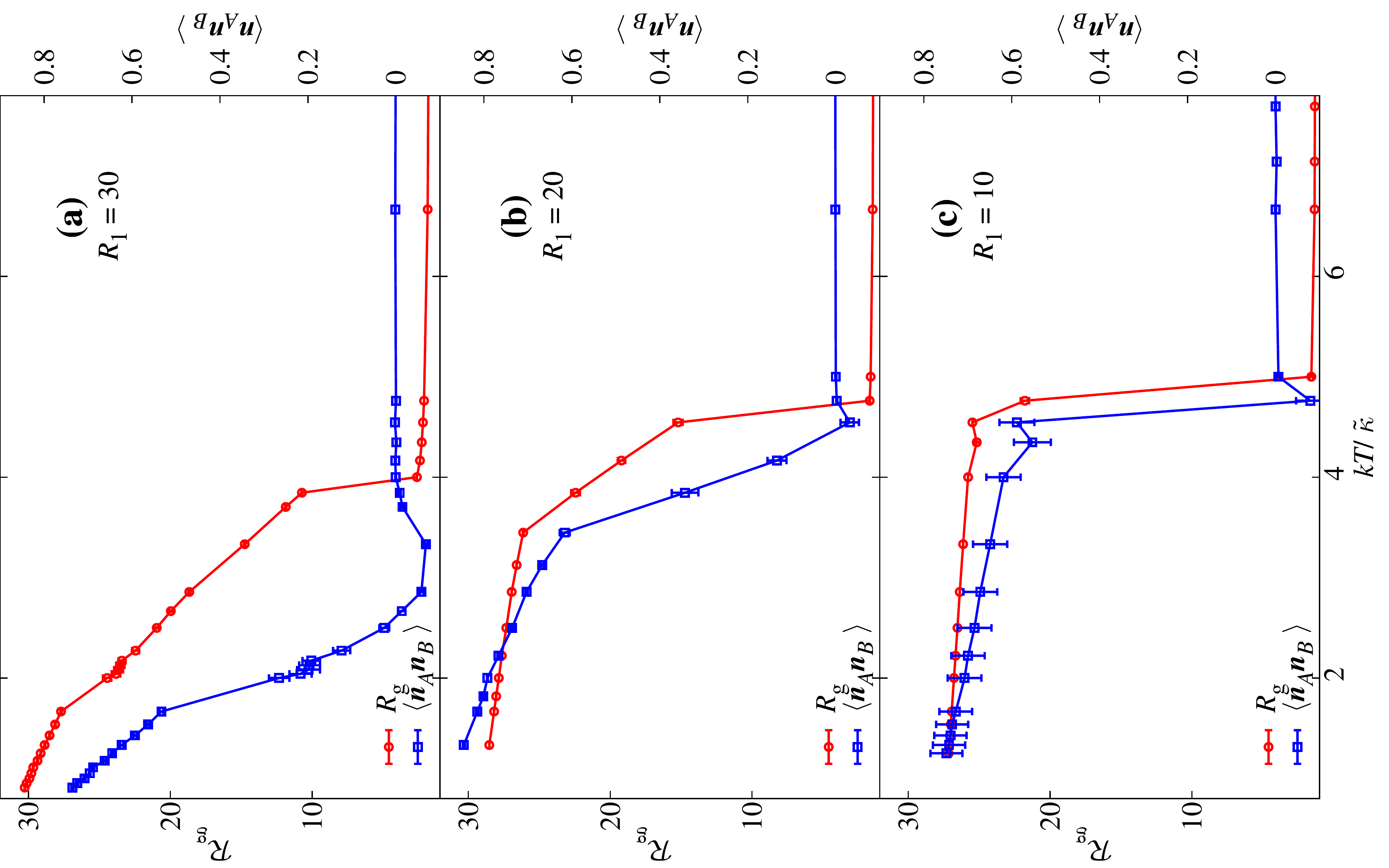}
\caption{Effect of the frame width on the folding regime.
We show the radius of gyration $\mathcal R_g$ (left axis, circles)
and the normal-normal correlations   $\langle \boldsymbol
n_A\cdot\boldsymbol n_B\rangle$ (right axis, squares) for
three different geometries. The outer radius is fixed at $R_2=R_1+W=42$,
with $R_1=30$ (a), $R_1=20$ (b) and $R_1=10$ (c).
The temperature interval between the point where $\langle \boldsymbol
n_A\cdot\boldsymbol n_B\rangle=0$ and the crumpling transition
is a visual representation of the shrinking folding regime
as $W$ increases.
\label{fig:size}}
\end{figure}

\begin{figure}
\includegraphics[width=\linewidth]{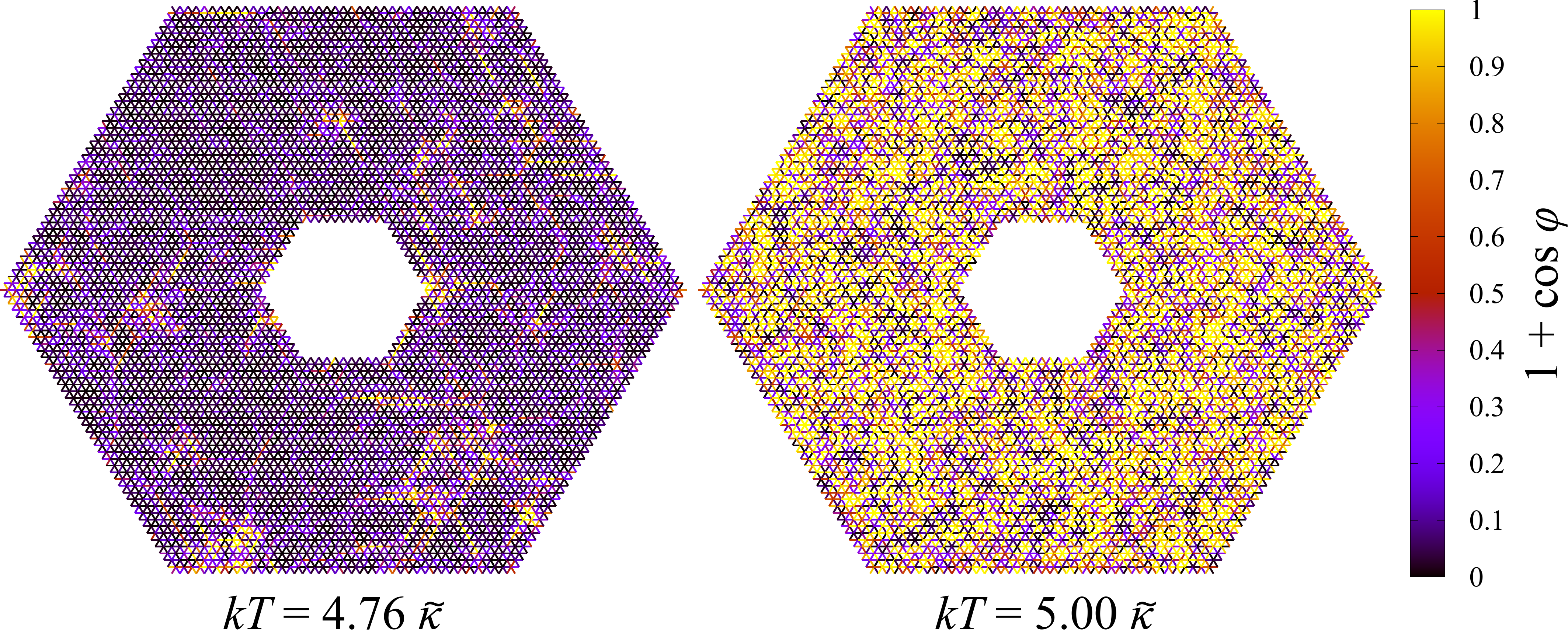}
\caption{
As in Figure~\ref{fig:heat}, but now for a much thicker frame ($W=32$). 
The system suddenly transitions from being almost flat with some 
localized larger fluctuations to being completely crumpled. We 
cannot identify the same individual folds as for the thin
frame.\label{fig:snapsR10}}
\end{figure}

We can test this idea of a temperature-dependent folding scale by plotting a
heat map of bending energy on the frame 
(Figure~\ref{fig:heat}).  For low temperatures, there is hardly any bending
energy in the system. As $T$ is increased, we find that almost all the bending
energy is concentrated in one or two radial segments, indicating well-defined
creases. The number of such folds increases with temperature until, at the
crumpling transition $kT_\text{c}\approx 4.0\tilde \kappa$, the whole system is
crumpled (i.e., folded at a microscopic scale).

\section{Normal-normal correlation function}\label{sec:corr}
We can make the analysis more quantitative by studying $\hat{\boldsymbol n}(r)$,
the normal to the sheet along the central hexagon of radius $R=(R_1+R_2)/2$, 
as a function of the arc length. Then we can compute
\begin{equation}\label{eq:corr}
C(r) = \sum_s \langle \hat{\boldsymbol n}(s)\cdot \hat{\boldsymbol n}(s+r)\rangle.
\end{equation}
This function gives us more complete information than the correlation between
particular pairs of points of the previous section and is also more
precise, since we can average over the whole curve of annular midpoints 
around the hexagon.

The result for $R_1=30$, $R_2=42$ is plotted in Figure~\ref{fig:corr}, where several
regimes can be identified. At the lowest temperatures, the system is deep
in the flat phase and $C(r)$ has a positive asymptote as $r\to\infty$. At the 
highest temperatures, in the crumpled phase, $C(r)$ is trivial: it is negative for
$r=2$ (opposite vertices of one dihedral) and zero beyond that since  the system
is folded at a microscopic scale. Qualitatively similar behavior was 
observed, e.g. \cite{Kantor1987}, for tethered surfaces with the geometry of
a parallelogram.  Our interest here
is the existence of an intermediate regime where $C(r)$ has negative values 
for a finite $r$ window, which shifts and shrinks as $T$ approaches $T_\text{c}$.
Just prior to this anomalous behavior, we find that
$C(r)$ is well described by an exponential decay $C(r)\simeq \exp(- r/\ell_\text{p})$, 
where $\ell_\text{p}$ is a persistence length (see, e.g.~\cite{Hoover1985,Kosmrlj2016,Yllanes2017}).
See the curves with $kT/\tilde\kappa \leq 2.00$ in Figure~\ref{fig:corr}.

Since the  $C(r)$ are very smooth functions and can be measured to high precision, 
we can easily find a minimum for each one in the folding regime. This minimum then
indicates the average distance between folds (or, alternatively, their average number) 
for each temperature  (Figure~\ref{fig:scale}).

\section{Role of the system geometry}\label{sec:geometry}
As we have mentioned, the $C(r)$ function has been extensively
studied for unperforated membranes.  In these cases, there is no
folding regime and the system suddenly switches between a flat
phase where $C(r)>0$ for all $r$ and an extremely compact
 crumpled phase where $C(r=2)<0$ and 
$C(r>2)=0$. This behavior arises because in unperforated membranes
the system presumably cannot 
find a privileged direction along which to fold in order to minimize
its bending energy.

We have tested these ideas by redoing our simulations for thicker
frames (smaller holes). In particular, keeping $R_2=42$, we have 
run simulations for $R_1=10,20$ ($W=22,32$, respectively, 
compared with the $W=12$ studied above). In this analysis, we have
found that for $W=22$ a folding regime can still be identified
(though it spans a much narrower temperature interval than for
$W=12$), while for $W=32$ the behavior is already qualitatively 
the same as for sheets without a hole~\footnote{And quantitatively 
very similar, with $kT_\text{c}$ approaching the value 
without a hole, although, unlike for our thin frames,
in this case finite-size effects in $kT_\text{c}$ are noticeable.}.

\begin{figure}
\includegraphics[height=\linewidth,angle=270]{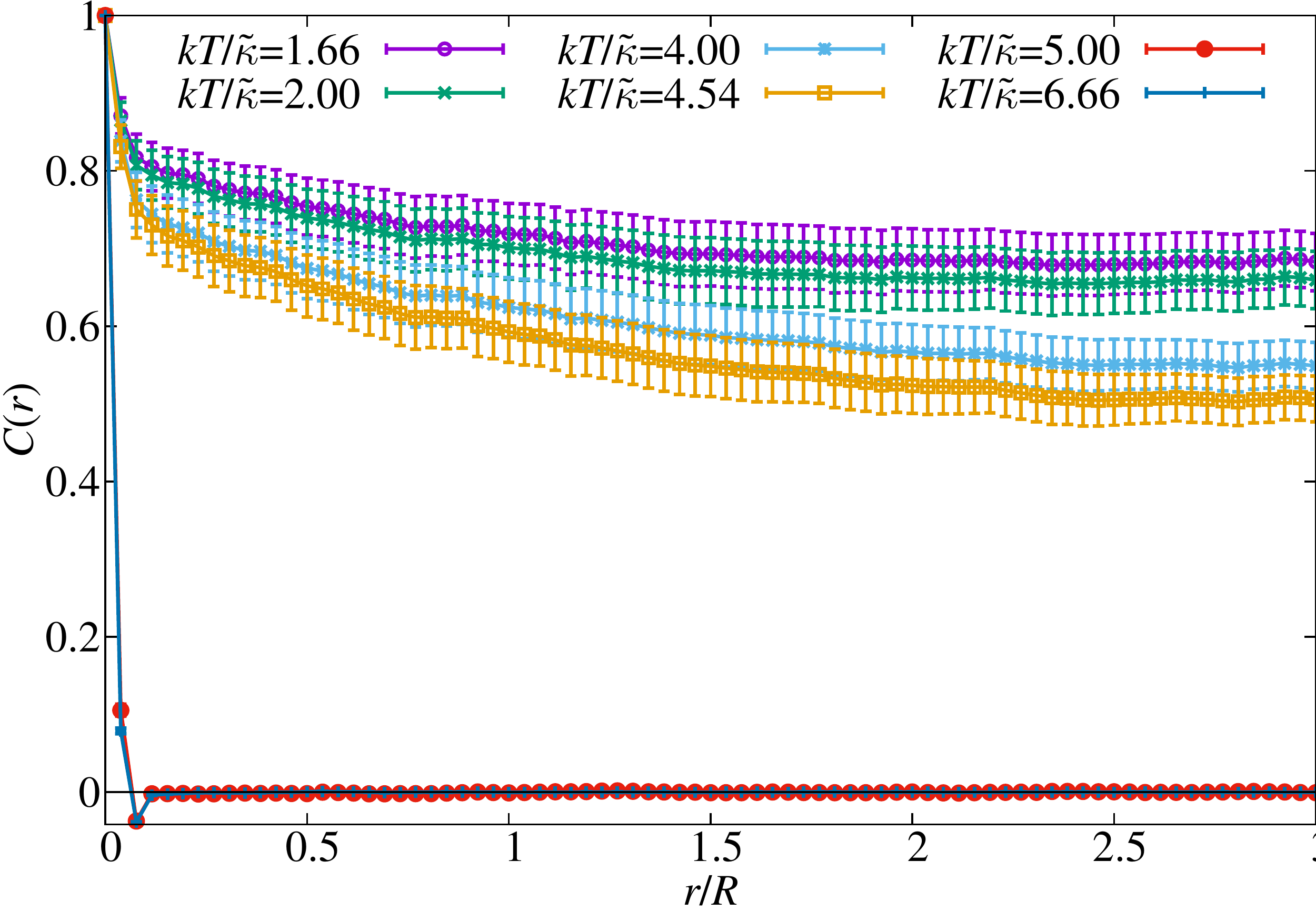}
\caption{Normal-normal correlation function as in Figure~\ref{fig:corr}, but
for a very thick frame ($R_1=10$, $R_2=42$).
Now we cannot identify an intermediate regime between
the flat and crumpled phases showing anticorrelation (folds) at finite distances.
\label{fig:corrR10}}
\end{figure}

This study is summarized in Figures~\ref{fig:size} to~\ref{fig:corrR10}. 
First, Figure~\ref{fig:size} plots the radius of gyration and \nAB\
for the three frame widths. As discussed in the previous sections,
the interval between the temperature at which $\nAB=0$ (appearance of very large folds)
and $T_\text{c}$ corresponds to the folding regime. This is very wide
for $W=12$, noticeably narrower for $W=22$ and has essentially disappeared
for $W=32$.

Moreover, if we consider (Figure~\ref{fig:snapsR10}) the same bending energy heat map
we used for our original frames, we see that even at a temperature just $5\%$ 
below $T_\text{c}$ no folds can be identified for our $W=32$ frame. The 
would-be folding 
regime has already shrunk beyond our ability to detect it. Finally, the picture
is confirmed by considering $C(r)$ again in Figure~\ref{fig:corrR10}, where
the correlation functions suddenly change from having a positive asymptotic value
(flat phase) to collapsing completely. This is the same pattern observed for 
membranes without holes in Ref.~\cite{Kantor1987}. 

\section{Conclusions}\label{sec:conclusions}
We have studied the equilibrium configurations of thermalized 
elastic frames as temperature is increased and found that, as the crumpling
transition is approached, most of the bending energy is concentrated 
on a growing number of origami-like folds. The scale of these folds
is a well-defined function of temperature.

The folding pathway explored here neglects distant
self-avoidance. In Ref. \cite{Yllanes2017}
we argued that the crumpling transition studied here for phantom frames
with large holes would persist even in the presence of self-avoidance in an 
appropriate thermodynamic limit. The non-trivial width-dependent scaling
of the thermally generated renormalized persistence length associated 
with our hexagonal frames
\begin{align}\label{eq:Lp}
\ell_\text{p} &= \frac{2W \kappa^\text{R}(W)}{kT}, &
\kappa^R(W) &= \kappa_0 \left(\frac{W}{\ell_\text{th}}\right)^\eta, &
\eta&\approx0.8,
\end{align}
suggests that the appropriate limit is $R,W\to\infty$ with fixed $W(W/\ell_\text{th})^\eta/R$
for frames with width $W$ and edge length $R$. Thus, we expect
that a sharp transition survives for hexagonal frames, where both a crumpled
and a flat phase would survive in a polymer-like large-size limit. In addition, we expect
that the origami-like folding pathway uncovered here will not be significantly affected 
by distant self-avoidance for most of the run-up to the crumpling transition from
low temperatures. Indeed, few self-intersections are evident in the partially 
crumpled images shown in Figs. \ref{fig:hexagono-crumpling} and \ref{fig:heat}. How
self-avoidance affects the folding at the transition itself when $kT\approx 4.0\tilde\kappa$
is an interesting subject for future investigation.

Some insight into the effect of self-avoidance on folding patterns without holes follows from the
study of Vliegenthart and Gompper of forced crumpling of self-avoiding elastic sheets 
at zero temperature \cite{Vliegenthart2006}. Here, a comparison of gradual forced
crumpling with and without self-avoidance eventually reveals a radial bias, just as
we find here with a single large hole. The patterns of wrinkles in this computer
experiment are qualitatively similar for both phantom and self-avoiding membranes.

\begin{acknowledgments}
This research was supported by the NSF through the DMREF grant DMR-1435794 and
DGE-1068780, as well as  by the Syracuse University Soft Matter Program.  Work
by DRN was also supported through the NSF DMREF program, via NSF grant
DMR-1435999 and via the Harvard Materials Science Research and Engineering
Center, through NSF grant DMR-1420570. The research of MJB was supported in
part by the National Science Foundation under Grant No. NSF PHY-1748958.  DY
thanks the KITP for hospitality during part  of this project and acknowledges
funding by  Ministerio de Econom\'ia, Industria y Competitividad (MINECO)
(Spain) through grants no. FIS2015-65078-C2 and PGC2018-094684-B-C21 (also partly funded by the EU
through the FEDER program) and the resources and assistance provided by
BIFI-ZCAM (Universidad de Zaragoza), where we carried out most of our
simulations on the Cierzo supercomputer.

\end{acknowledgments}

\appendix
\section{The normal-normal correlation in Fourier space}\label{sec:fourier}
\begin{figure}
\includegraphics[height=\linewidth,angle=270]{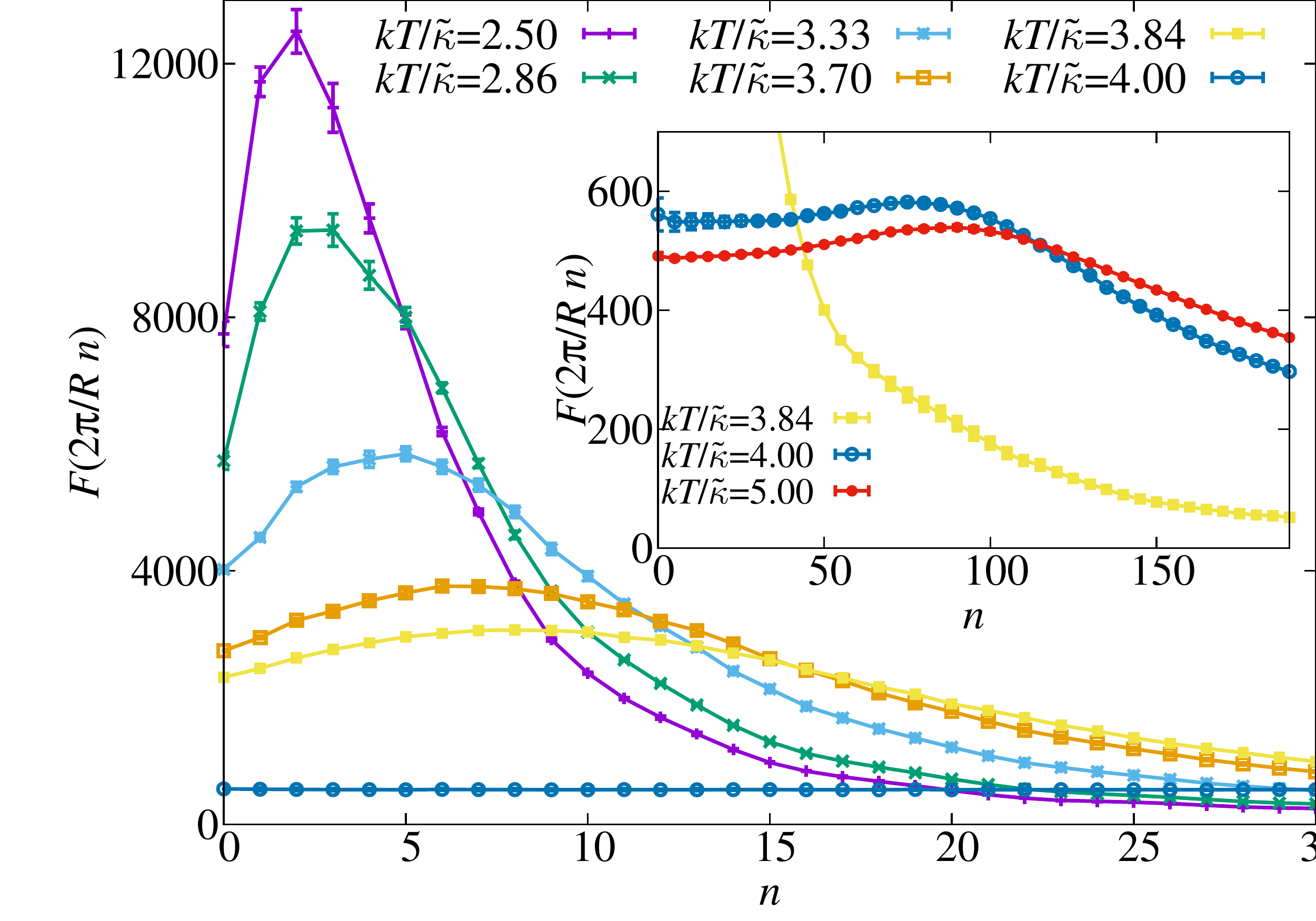}
\caption{
We plot the full normal-normal correlation function in Fourier space, defined
by Eq. \eqref{eq:F}, for 
several temperatures. Since in our discretized system, the only available
spatial wavevectors are $k_n = n (2\pi/R)$, in this case we show the results
for a larger system than in previous figures ($R_1=70, W=12, R=76$).
There is a well-defined maximum for each temperature in the extended phase
and a dramatic transition in the shape of the function once
we reach the crumpling temperature $kT_\text{c}\approx 4.0\kappa$. The \emph{inset}
shows a closeup of the structure function as we move deep into the crumpled phase (note the scales
of the axes). 
\label{fig:F}}
\end{figure}

\begin{figure}
\includegraphics[height=\linewidth,angle=270]{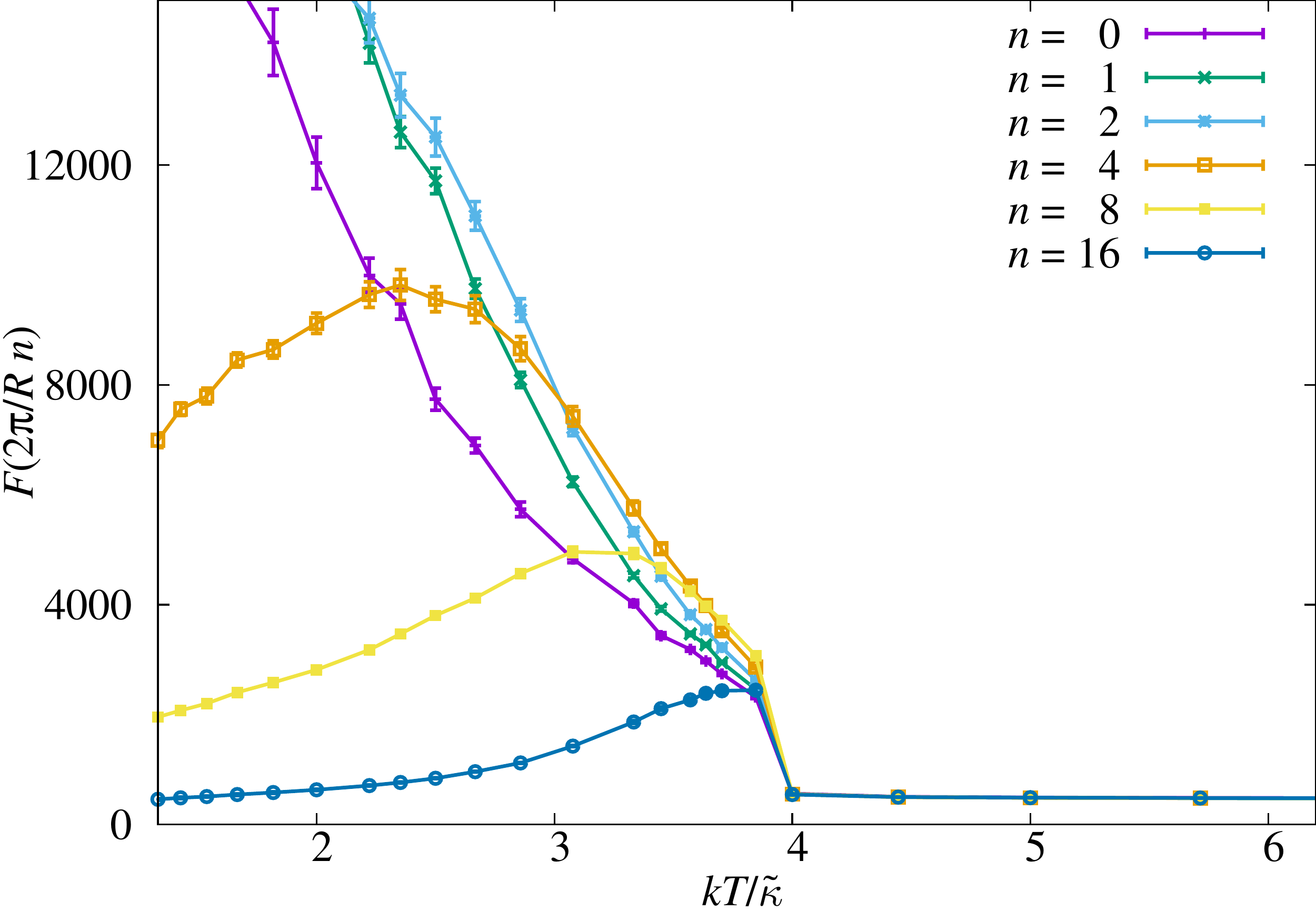}
\caption{The zeroth component of the correlation 
function \eqref{eq:F}, $F(k=0)$, is also a good order parameter for the crumpling transition. Here, we plot
this and other components of $F$ as a function of temperature.
\label{fig:F0}}
\end{figure}

In the main part of this paper we have  worked with the correlation 
function in real (position) space. However, its Fourier transform
\begin{equation}\label{eq:F}
F(k) = \biggl\langle \biggl|\sum_r C(r) \mathrm{e}^{\mathrm{i} k r}\biggr|^2\biggr\rangle
\end{equation} 
provides an interesting complementary picture. Now the folding regime 
leads to a \emph{maximum} in  $F(k)$ that shifts to smaller wavenumbers
as $T$ increases (Figure~\ref{fig:F}). In addition, considering $F(k)$ 
as a function of temperature offers an alternative way of finding
the crumpling point (Figure~\ref{fig:F0}). 

\section{The crumpling process}
\begin{figure*}[p]
\centering
\includegraphics[height=.6\linewidth,angle=270]{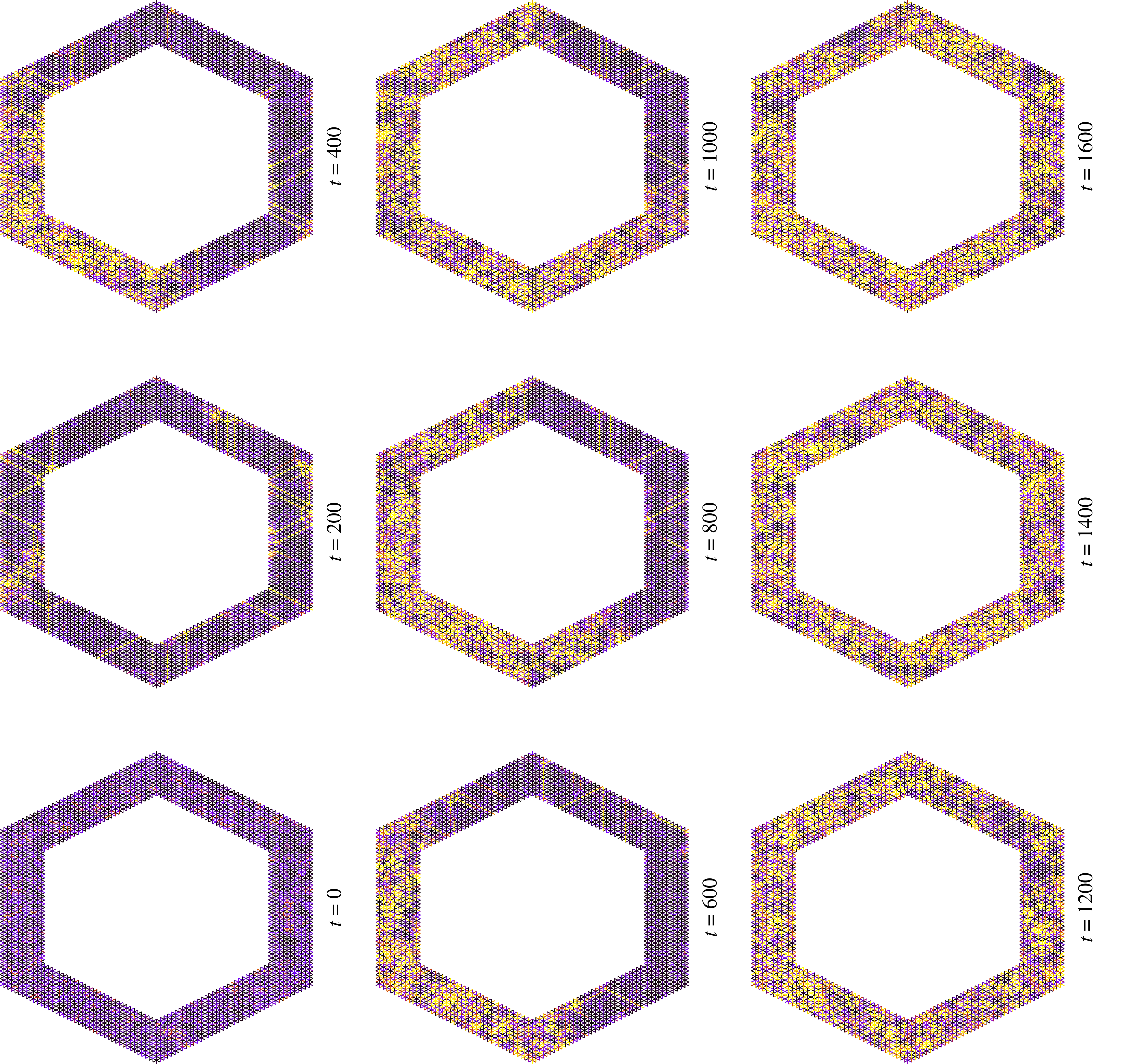}
\caption{Time evolution of the bending energy
with $R_1=30$, $R_2=42$ at the crumpling transition
$kT=4.00\tilde \kappa\approx k T_\text{c}$ (times are 
in units of $10^4$ MD steps).
Note that linear folds appear in portions of the frame 
at intermediate times along this pathway.
\label{fig:colapsoR30}}

\includegraphics[height=.6\linewidth,angle=270]{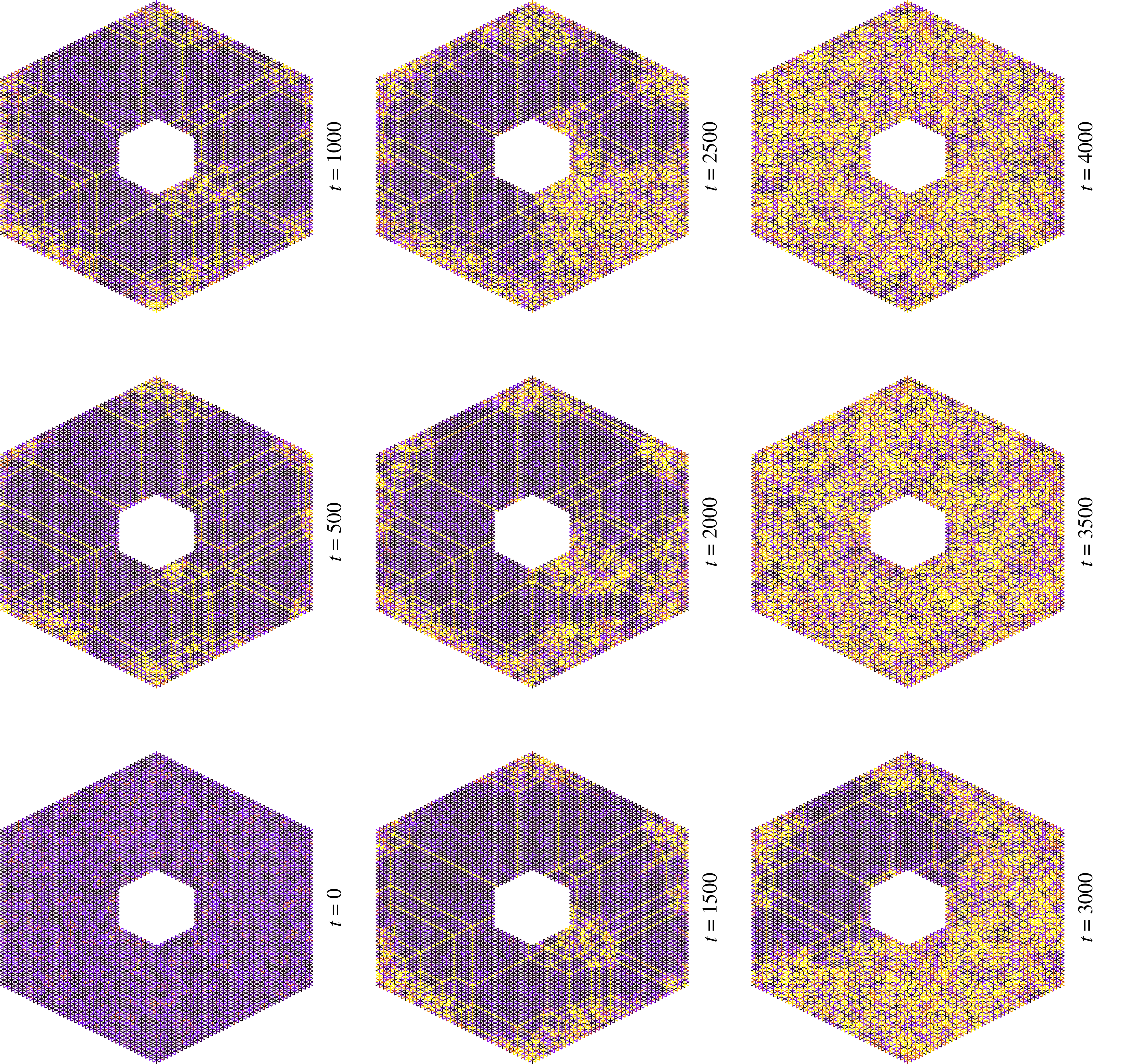}
\caption{Time evolution of the crumpling transition of a 
frame with a smaller hole, $R_1=10$, $R_2=42$. Here
the crumpling transition occurs at a higher temperature
and we set 
$kT=5.00\tilde \kappa \approx T_\text{c}$.\label{fig:colapsoR10}}
\end{figure*}

In the main text of the paper,
we considered the folding pathways as the temperature 
was increased up to the crumpling transition, which
we studied through the analysis of configurations extracted
from the equilibrium evolution of the system. We can obtain a complementary
picture by studying the non-equilibrium evolution of a system at $T_\text{c}$
which started from a completely flat configuration.
We have done this in Figures~\ref{fig:colapsoR30} and~\ref{fig:colapsoR10}
for our systems with $R_2=42$ and $R_1=10,30$ (each at their corresponding
$T_\text{c}$). For the latter, individual folds quickly appear at several
points along the frame and serve to nucleate crumpled regions. It takes longer
for the thicker frame to crumple, since the system cannot create linear folds
that easily. In both cases, once the system has crumpled completely, the microscopic
fold patterns correspond to those of the planar folding problem, as in~\cite{DiFrancesco1994}.

\bibliographystyle{apsrev4-1}
\bibliography{biblio-soft}

\end{document}